\documentclass[aps,prd,nofootinbib,superscriptaddress,preprint,tightenlines]{revtex4-1}
\usepackage{amssymb, bbm, amsmath, amsfonts, amsthm, amsbsy, mathtools, mathrsfs, tensor}
\usepackage{enumerate}
\usepackage{graphicx}
\graphicspath{ {Images/} }
\usepackage{color}
\usepackage{hyperref}
\usepackage[marginal, multiple]{footmisc}
\usepackage{cleveref}

\newcommand {\pd} [2] {\frac {\partial #1} {\partial #2}}

\newcommand {\intf} [2] {\int_{#1}^{#2}}

\newcommand{\inp} [1] {\left( #1 \right)}

\newcommand{\insb} [1] {\left[ #1 \right]}

\def \vphi {\varphi\,}

\def \lie {\pounds}

\def\R{\mathbb{R}}

\theoremstyle{definition}

\theoremstyle{remark}

\begin{document}
	\title{Black Hole Memory}
	
	\author{Adel A. Rahman}
	\email{arahman@stanford.edu}
	\affiliation{\small Department of Physics and the Enrico Fermi Institute, The University of Chicago,\\ Chicago, Illinois 60637, USA \vspace{1mm}}
	\affiliation{\small Department of Physics, Stanford University\\
		Stanford, California 94305, USA \vspace{1mm}}
	\author{Robert M. Wald}
	\email{rmwa@uchicago.edu}
	\affiliation{\small Department of Physics and the Enrico Fermi Institute, The University of Chicago,\\ Chicago, Illinois 60637, USA \vspace{1mm}}
	
	\begin{abstract}
	\quad

The memory effect at null infinity, $\mathcal{I}^+$, can be defined in terms of the permanent relative displacement of test particles (at leading order in $1/r$) resulting from the passage of a burst of gravitational radiation. In $D=4$ spacetime dimensions, the memory effect can be characterized by the supertranslation relating the ``good cuts'' of $\mathcal{I}^+$ in the stationary eras at early and late retarded times. It also can be characterized in terms of charges and fluxes associated with supertranslations.
Black hole event horizons are in many ways analogous to $\mathcal{I}^+$. We consider here analogous definitions of memory for a black hole, assuming that the black hole is approximately stationary at early and late
advanced times, so that its event horizon is described by a Killing horizon (assumed nonextremal) at early and late times. We give prescriptions for defining preferred foliations of nonextremal Killing horizons. We give a definition of the memory tensor for a black hole in terms of the ``permanent relative displacement'' of the null geodesic generators of the event horizon between the early and late time stationary eras. We show that preferred foliations of the event horizon in the early and late time eras are related by a Chandrasekaran-Flanagan-Prabhu (CFP) supertranslation.
However, we find that the memory tensor for a black hole horizon does not appear to be related to the CFP symmetries or their charges and fluxes in a manner similar to that occurring at $\mathcal{I}^+$.

	\end{abstract}
	
	\maketitle 
	
\section{Introduction} 
	
A gravitational wave in an asymptotically flat spacetime passing through a system of test particles near null infinity, $\mathcal{I}^+$, will induce oscillations in the relative positions of these test particles. It has long been known that, after the wave has passed by, the relative positions of the particles need not return to their original values \cite{LinMem,Christodoulou}. This fact is known as the \emph{gravitational memory effect}. The memory effect has received considerable interest in recent years due to its connections with asymptotic symmetries, conservation laws, and infrared divergences \cite{StromingerBook,SatishWald,IRReview}.

Memory at null infinity can be characterized in a number of different ways. As we shall review in \cref{memnullinf}, memory at null infinity is defined in terms of the leading order in $1/r$ change in the relative displacement of test particles (initially at rest) between stationary eras. In $D=4$ spacetime dimensions, the memory tensor, $\Delta_{AB}$, defined in this way is given by a formula that expresses it in terms of a supertranslation, $T$, \cite{StromingerZ,HIWmem} (see \cref{GWMemasST} below). The supertranslation, $T$, also describes the relationship between the ``good cuts'' \cite{GoodCuts} of $\mathcal{I}^+$ in the initial and final stationary eras, where the good cuts are those for which the shear of the ``ingoing'' (i.e., transverse to $\mathcal{I}^+$) orthogonal null geodesics vanishes at $\mathcal{I}^+$ in the unphysical spacetime\footnote{In the physical spacetime, this corresponds to the vanishing of the shear at order $1/r^2$.}. Thus, memory at $\mathcal{I}^+$ can also be characterized in terms of the asymptotic symmetry that relates the good cuts of $\mathcal{I}^+$ in the stationary eras at early and late times. In addition, memory at $\mathcal{I}^+$ can also be characterized in the following way: Asymptotic symmetries at $\mathcal{I}^+$ have associated charges and fluxes \cite{WZ}. The integrated flux associated with a supertranslation has a contribution arising from the Bondi flux (the ``hard'' integrated flux), but it also has a contribution from the memory tensor (the ``soft'' integrated flux); see, e.g., section F1 of \cite{SatishWald}. Thus, the memory tensor characterizes the ``soft'' contribution to the integrated flux associated with supertranslation charges.

Black hole horizons have many features in common with $\mathcal{I}^+$. Both, of course, are null boundaries of the domain of outer communications of the black hole. The event horizon of a stationary black hole is a Killing horizon, and thus its null geodesic generators have vanishing expansion and shear, similar to $\mathcal{I}^+$. As shown in \cite{HW}, for perturbations of a stationary black hole, the canonical energy flux through the horizon is given by a formula that exactly mirrors the formula for Bondi energy flux through $\mathcal{I}^+$, with the perturbed shear of the horizon playing the role of Bondi news at $\mathcal{I}^+$. These close analogies suggest that there might be a notion of memory for black hole event horizons that has similar features to memory at $\mathcal{I}^+$. A notion of black hole memory and its properties could potentially be of interest for the investigation of classical and quantum aspects of black holes, particularly in view of the suggestion of \cite{HPS} that ``soft hair'' on black holes could play a
role in the black hole information issue.

In this paper, we will consider notions of black hole memory\footnote{A notion of black hole memory has been previously considered by Donnay, Giribet, Gonz\'alez, and Puhm \cite{BHME} from a perspective quite different from ours. The idea that changes in horizon foliation may play a role in obtaining a notion of black hole memory was previously suggested in \cite{MWZ}.} and their properties. We consider a black hole that becomes (approximately) stationary at early and late retarded times---so that its event horizon, $\mathcal{H}^+$, is well approximated
by a Killing horizon (assumed to be nonextremal) in these eras---and we investigate the extent to which black hole memory can be defined and characterized in a manner analogous to memory at null infinity. We find the following: (1) One can define a memory tensor, $\Delta_{AB}$, at the event horizon, $\mathcal{H}^+$, of a black hole analogous to the memory tensor at $\mathcal{I}^+$. The black hole memory tensor directly describes the net change in the relative displacement of the null generators of the horizon between the early and late time stationary eras. (2) We show that natural analogs of ``good cuts'' can be defined for any nonextremal Killing horizon. Indeed, we find two inequivalent prescriptions for defining good cuts. Using either of these prescriptions, we show that the early time good cuts are related to the late time good cuts by a Chandrasekaran-Flanagan-Prabhu  (CFP) \cite{CFP} supertranslation, $T$. However, unlike the situation at $\mathcal{I}^+$, there does not appear to be any relation between $T$ and the memory tensor $\Delta_{AB}$. There also does not appear to be any relationship between the charges and fluxes associated with CFP supertranslations and the memory tensor $\Delta_{AB}$.

Thus, we find that many of the facets of memory at $\mathcal{I}^+$ have close analogs at the event horizon of a black hole. However, the relationships between these facets that hold at $\mathcal{I}^+$ do not appear to hold at the horizon of a black hole. It is not difficult to identify the mathematical reason why this is so. A gravitational wave at $\mathcal{I}^+$ does not affect the ``zeroth order'' structure of $\mathcal{I}^+$, i.e., it affects the metric only at ``radiative order,'' $1/r^{(D/2 - 1)}$, and faster fall-off, where $D$ is the spacetime dimension. In particular, the geometry of $\mathcal{I}^+$ itself is unchanged by the passage of a gravitational wave. By contrast, a gravitational wave passing through the event horizon of a black hole has a ``zeroth order'' effect on the geometry of the horizon. Black hole memory affects the horizon itself rather than being a ``perturbative effect'' away from the horizon. For this reason, relationships between aspects of memory at a black hole horizon have a very different mathematical structure from those at $\mathcal{I}^+$.

In \cref{memnullinf} we briefly review the notion and properties of memory at $\mathcal{I}^+$, focusing attention on $D=4$ spacetime dimensions.
In \cref{RevNull}, we briefly review properties of null hypersurfaces and the construction of Gaussian null coordinates. We also briefly review the notion of supertranslation symmetries of null hypersurfaces recently introduced by Chandrasekaran, Flanagan, and Prabhu  (CFP) \cite{CFP}. 

Preferred foliations of Killing horizons by ``good cuts'' are analyzed in \cref{KHFol}. We cannot define good cuts of a Killing horizon by requiring the vanishing of the shear of the orthogonal null geodesics transverse to the cut, since, in general, no such cuts will exist. Nevertheless, we will show that any nonextremal Killing horizon admits a unique foliation by cuts whose transverse expansion is uniform over the cut. We also show that nonextremal Killing horizons in spacetimes admitting a $t$--$\phi$ reflection isometry also possesses a unique foliation associated with this reflection symmetry. In addition, Ashtekar, Beetle, and Lewandowski (ABL) \cite{ABL1, ABL2} have also given a prescription for obtaining a unique foliation of a nonextremal Killing horizon\footnote{The prescription was given in \cite{ABDFKLW,ABL1,ABL2} in the slightly more general context of a nonextremal weakly isolated horizon}. We show in \cref{KHFol} that (where it is defined) the $t$--$\phi$ reflection foliation agrees with the ABL foliation, but that these foliations do not, in general, agree with the uniform expansion foliation. Thus, we have two inequivalent prescriptions for defining a preferred foliation of a Killing horizon by ``good cuts.''

Black hole memory is then analyzed in \cref{BHMem}. We define a memory tensor for black holes in \cref{bhmemten}. We show in \cref{BHMemProp} that the early and late time preferred foliations (using either of the above prescriptions) are related by a CFP supertranslation.

We conclude in \cref{Disc} with a brief discussion of the differences between black hole memory and memory at null infinity.

For the most part, our conventions and notations follow those of \cite{Wald}. In particular, we use the ``mostly pluses" convention for the signature $(-++\dots +)$ of the metric, work in geometrized units with $c = G_N = 1$, and employ abstract index notation (see section 2.4 of \cite{Wald}) for tensor fields on spacetime and its submanifolds. Greek indices (e.g. $\mu, \nu, \rho, \dots $) will be used to denote coordinates $x^{\mu}$ on spacetime and components of spacetime tensors in the corresponding coordinate bases. Tensors on spacetime will be denoted by lowercase Latin indices from the beginning of the alphabet (e.g. $a, b, c, \dots$). Tensors on codimension-one hypersurfaces will be denoted by lowercase Latin indices from the middle of the alphabet (e.g. $i, j, k, \dots$). Tensors on surfaces of codimension two, such as cross-sections of a null hypersurface, will be denoted by uppercase early Latin indices (e.g. $A, B, C, \dots$). In order to avoid using an additional alphabet, we will also use capital Latin indices from the beginning of the alphabet to denote coordinates $x^A$ on codimension-two submanifolds of $\mathcal{M}$ and components of tensors in the corresponding coordinate bases. We will also use uppercase early Latin indices to denote tensors in the tensor algebra $\mathsf{W}$ of a null hypersurface introduced in \cref{Basic}. $\Phi^*$ and $\Phi_*$ will denote, respectively, pushforward and pullback by a smooth map $\Phi$, and $\lie_X$ will denote the Lie derivative with respect to the vector field $X^a$. The symbol $\widehat{=}$ will denote equality when both sides are restricted to a given null hypersurface.

Our discussion and results on null hypersurfaces, Killing horizon foliations, and black hole memory apply to all spacetime dimensions $D \geq 4$. However, for simplicity, we will restrict the discussion of results on memory at null infinity that we review in \cref{memnullinf} to the $4$-dimensional case.

\section{Memory at Null Infinity}
	\label{memnullinf}
Asymptotic flatness at null infinity in even dimensional spacetimes can be defined in terms of the ability to do a ``conformal compactification'' of spacetime in null directions. More precisely, a spacetime $(\mathcal{M}, g_{ab})$ is said to be asymptotically flat at null infinity if there exists an ``unphysical" spacetime $(\tilde{\mathcal{M}}, \tilde{g}_{ab})$ and an embedding $i: \mathcal{M} \hookrightarrow \tilde{\mathcal{M}}$, with $\tilde{g}_{ab} = \Omega^2(i^*g)_{ab}$ (for some highly non-unique ``conformal factor" $\Omega \in C^{\infty}(\tilde{\mathcal{M}})$) satisfying the properties discussed in e.g. \cite{Wald, HIWmem, Geroch}. Future null infinity, $\mathcal{I}^+$, is (a component of) the smooth hypersurface $\Omega = 0$ (the \emph{conformal boundary} of $\mathcal{M}$) in the unphysical spacetime. We may choose $\Omega$ so that $\mathcal{I}^+$ has vanishing expansion and shear in the unphysical spacetime. Using $\Omega$ as a coordinate, we may put the unphysical metric near $\mathcal{I}^+$ in the form\footnote{This form may be obtained by using Gaussian null coordinates (see \cref{Basic} below) in the unphysical spacetime.}
	\begin{equation}
	d\tilde{s}^2 = 2\,dud\Omega -2\,\Omega\,\tilde{\alpha}\,du^2-2\,\Omega\,\tilde{\beta}_A\,dudx^A + \tilde{\gamma}_{AB}\,dx^Adx^B
	\label{ConfFrame}
	\end{equation}
where 
	\begin{equation}
\Omega^{-1}\tilde{\alpha}|_{\mathcal{I}^+} = 1/2 \, , \qquad \tilde{\gamma}_{AB}|_{\mathcal{I}^+} = s_{AB}
	\end{equation}
with $s_{AB}$ the ``round sphere" metric on $\mathbb{S}^{2}$. Using Einstein's equation, we then have $\beta_A|_{\mathcal{I}^+} = 0$ \cite{HIW}. ``Bondi coordinates" $(u,r, x^A)$ in the physical spacetime can be obtained by replacing the coordinate $\Omega$ with $r \coloneqq 1/\Omega$, i.e.
	\begin{equation}
	ds^2 = -2\,dudr -2r\tilde{\alpha}\,du^2 -2r\tilde{\beta}_A\,dudx^A + r^2\,\tilde{\gamma}_{AB}\,dx^Adx^B
	\label{scricoord}
	\end{equation}
where $\tilde{\alpha}$, $\tilde{\beta}_A$, and $\tilde{\gamma}_{AB}$ can be (asymptotically) expanded in powers of $1/r$ (since they are smooth in $\Omega$). 
A smooth conformal compactification cannot be done for odd-dimensional radiating spacetimes \cite{HWOdd}, but asymptotic flatness at null infinity can be defined in odd dimensions in terms of the ability to perform a suitable $1/r$ expansion of the metric \cite{SatishWald}. In the remainder of this section, we will restrict consideration to $D=4$ spacetime dimensions. A discussion of all aspects of memory at null infinity in higher dimensions can be found in \cite{SatishWald}.

Asymptotic symmetries at $\mathcal{I}^+$ are defined to be diffeomorphisms which preserve the asymptotic form of the metric modulo the diffeomorphisms that are degeneracies of the symplectic form \cite{WZ}. The group of asymptotic symmetries at $\mathcal{I}^+$ in $D=4$ spacetime dimensions is known as the \emph{BMS Group} \cite{BBM, Sachs} and has the group structure
	\begin{equation}
	\mathscr{G}_{\mathrm{BMS}} = SO(3,1) \ltimes \mathscr{S}_{\mathrm{BMS}}
	\end{equation} 
where the group of \emph{BMS supertranslations} $\mathscr{S}_{\mathrm{BMS}}$ acts on $\mathcal{I}^+$ as 
	\begin{equation}
	u \mapsto u + T(x^A)
	\label{superT}
	\end{equation}
The $4$-dimensional subgroup $\mathscr{T}$ of $\mathscr{S}_{\text{BMS}}$ with $T$ equal to some linear combination of the $\ell = 0,1$ spherical harmonics is simply the ordinary translation group, while the remaining elements of $\mathscr{S}_{\text{BMS}}$ are referred to as the nontrivial BMS supertranslations.

The \emph{Bondi news tensor}, $N_{AB}$, at $\mathcal{I}^+$ is given by
	\begin{equation}
	N_{AB} \coloneqq \left({s_A}^C {s_B}^D - \frac{1}{2} s_{AB}\, s^{CD} \right) \partial_u \tilde{\gamma}^{(1)}_{CD}
	\label{News}
	\end{equation} 
where $\tilde{\gamma}^{(1)}_{CD}$ denotes the order $1/r$ part of $\tilde{\gamma}_{CD}$.
As originally shown in \cite{GoodCuts}, if the Bondi news tensor vanishes---i.e, in the absence of outgoing radiative modes---one can find cross-sections of $\mathcal{I}^+$ that have the property that, in the physical spacetime, the shear of the ingoing orthogonal null geodesics (i.e., the orthogonal null geodesics transverse to $\mathcal{I}^+$ in the unphysical spacetime) vanishes at order $1/r^2$ (or, equivalently, that the shear of the orthogonal null geodesics transverse to $\mathcal{I}^+$ vanishes at $\mathcal{I}^+$ when computed using the unphysical spacetime metric). Such cross-sections are called ``good cuts'' and are unique up to ordinary translations. In an era where the Bondi news vanishes, the good cuts can be used to pick out a preferred Poincar\'e subgroup of $\mathscr{G}_{\text{BMS}}$.
	
The memory effect at $\mathcal{I}^+$ refers to the net relative displacement attained by a collection of inertial test masses near $\mathcal{I}^+$ in an asymptotically flat spacetime after the passage of a burst of gravitational radiation \cite{LinMem, Christodoulou}. More precisely, suppose that an asymptotically flat spacetime $(\mathcal{M}, g_{ab})$ is stationary at order $1/r$ at early times, $u < u_0$, and at late times, $u > u_1$, so that, in particular, the Bondi news vanishes in these eras. Suppose that two nearby inertial (i.e. geodesic) test masses at large $r$ have worldlines initially tangent to $(\partial/\partial u)^a$ and have initial angular deviation vector $\xi^A_0$. During the nonstationary era, the time-evolved deviation vector, $\xi^A(u)$, will differ from the parallel transport of $\xi^A_0$ at order $1/r$. In the late time stationary era $u > u_1$, the deviation vector will again become time independent at order $1/r$, but there will remain an $O(1/r)$ difference between $\xi^A$ and the parallel transport of $\xi^A_0$. The {\em memory tensor}, $\Delta_{AB}$, is defined by
		\begin{equation}
		\xi^{(1) A} = \Delta\indices{^A_B}\,\xi^B_0
		\label{DispTensorDef}
		\end{equation}
where the $\xi^{(1) A}$ denotes the difference at order $1/r$ between $\xi^A$ and the parallel transport of $\xi^A_0$ in the late time stationary era.

The memory tensor can be computed by integrating the geodesic deviation equation through the radiating era. One finds that \cite{Christodoulou}
	\begin{equation}
	\Delta_{AB} = \frac{1}{2}\,\tilde{\gamma}^{(1)}_{AB} \Big|^{u = \infty}_{u =- \infty} 
	\label{MemTensor}
	\end{equation} 
where any coordinates compatible with a suitable $1/r$ expansion can be used to compute the right side of \cref{MemTensor} \cite{SatishWald}. In higher dimensions, a direct generalization of this formula holds if we use ``comoving coordinates'' near null infinity for the family of geodesics initially tangent to $(\partial/\partial u)^a$, i.e. coordinates in which $(r, x^A)$ are constant along these geodesics (see section IIIE of \cite{SatishWald}).

It can further be shown that---in the absence of magnetic parity memory \cite{SatishWald}---$\Delta_{AB}$ takes the form \cite{StromingerZ, HIWmem}
	\begin{equation}
	\Delta_{AB} = -\inp{{D}_A {D}_B-\frac{1}{2} s_{AB}\,{D}^C{D}_C}T \qquad 
	\label{GWMemasST}
	\end{equation}
where $D_A$ is the derivative operator on $\mathbb{S}^{2}$ with round metric $s_{AB}$, and $T$ is a function on $\mathbb{S}^{2}$. We can naturally identify $T$ with a supertranslation via \cref{superT}.

As mentioned above, in the stationary eras $u < u_0$ and $u > u_1$ we may define ``good cuts'' of $\mathcal{I}^+$, which are unique up to translations in these eras. However, the good cuts in the region $u < u_0$ will not, in general, be related to the good cuts of the region $u > u_1$ by 
an ordinary translation alone. Indeed, they are related by the same supertranslation that appears in \cref{GWMemasST}. To see this, consider Bondi coordinates \cref{scricoord}
chosen so that for $u < u_0$ the cross-sections of constant $u$ are good cuts of $\mathcal{I}^+$. Under a supertranslation $u \mapsto u + f$, we have 
		\begin{equation}
\tilde{\gamma}^{(1)}_{AB} \mapsto \tilde{\gamma}^{(1)}_{AB} + 2D_AD_Bf
		\end{equation}
so that $\tilde{\sigma}_{AB}^{(2)}$, the order $1/r^2$ part of the shear of the ingoing orthogonal null geodesics, transforms as,
		\begin{equation}
		\tilde{\sigma}^{(2)}_{AB} = \frac{1}{2}\inp{\tilde{\gamma}^{(1)}_{AB}-\frac{1}{2}s_{AB}\,s^{CD}\tilde{\gamma}^{(1)}_{CD}} \mapsto \tilde{\sigma}'^{(2)}_{AB} = \tilde{\sigma}^{(2)}_{AB} + \inp{D_AD_B - \frac{1}{2}s_{AB}\,D^CD_C}f
		\label{sigmachange}
		\end{equation}
Since $\tilde{\sigma}^{(2)}_{AB} = 0$ at early times, we have by \cref{MemTensor} that
		\begin{equation}
		\Delta_{AB} = \tilde{\sigma}^{(2)}_{AB}\big|_{\text{late times}} 
		\label{DispAsShear}
		\end{equation}
The late time good cuts, on the other hand, will be given by the surfaces of constant ${u' = u + f(x^A)}$, where, via \cref{sigmachange} and \cref{DispAsShear},
		\begin{equation}
		0 = \tilde{\sigma}'^{(2)}_{AB} = \Delta_{AB} + \inp{D_AD_B - \frac{1}{2}s_{AB}\,D^CD_C}f
		\end{equation}
Thus, the supertranslation $f$ relating the early and late time good cuts of $\mathcal{I}^+$ coincides with the supertranslation $T$ appearing in \cref{GWMemasST}.

Finally, we note that associated with any asymptotic symmetry at $\mathcal{I}^+$ is a corresponding charge and flux \cite{WZ}. The integrated flux associated with a supertranslation $u \mapsto u + \alpha(x^A)$ is given by (see eq.(218) of \cite{SatishWald})
	\begin{equation}
	\int_{\mathcal{I}^+} {\mathcal F}_{(\alpha)} = -\frac{1}{32 \pi} \int_{\mathcal{I}^+} \alpha\,N_{AB} N^{AB} + \frac{1}{8 \pi} \int_{\mathbb{S}^{2}} \alpha \,D^A D^B \Delta_{AB}
	\label{intflux}
	\end{equation}
This gives another characterization of memory as a contribution to the integrated fluxes associated with supertranslation charges. This contribution is usually referred to as the ``soft'' part of the integrated flux. Note that the soft integrated flux associated with $\alpha = T$ itself is
	\begin{equation}
	\int_{\mathcal{I}^+} {\mathcal F}^{\rm soft}_{(T)} = - \frac{1}{8 \pi} \int_{\mathbb{S}^{2}} \Delta^{AB} \Delta_{AB}
	\label{softflux}
	\end{equation}

	\section{Null Hypersurfaces} 
	\label{RevNull}

In this section, we will review some basic definitions and constructions for null hypersurfaces in $D$-dimensional spacetimes, as well as the recent work of Chandrasekaran, Flanagan, and Prabhu \cite{CFP} on the symmetry group of a null surface. 
	
	\subsection{Basic Notions and Constructions}
	\label{Basic}
	
	Let $(\mathcal{M}, g_{ab})$ be a $D$-dimensional, time oriented Lorentzian spacetime 
	and let $\mathcal{N}$ be a null hypersurface in $\mathcal{M}$---i.e. a smooth codimension-one submanifold of $\mathcal{M}$ whose normal 
is a smooth future-directed null vector field, $k^a$, which is defined on $\mathcal{N}$ up to scaling by a strictly positive function $k^a \to e^{f}k^a$, $f \in C^{\infty}(\mathcal{N})$. 
The integral curves of $k^a$ are null geodesics (which are not necessarily affinely parameterized and may be incomplete), and we assume that each point in $\mathcal{N}$ lies on a unique such integral curve, or ``null generator". 
We further assume that $\mathcal{N}$ is in fact diffeomorphic to a product $\mathcal{N} \simeq \bar{\mathcal{S}}\times\R$, where $\bar{\mathcal{S}}$ is the manifold of integral curves of $k^a$.
	
$\mathcal{N}$ inherits from $(\mathcal{M}, g_{ab})$ a (degenerate) metric $q_{ij}$ and a $q_{ij}$ compatible connection $\mathcal{D}_i$. For a given choice of $k^a$, we also obtain a volume form $\boldsymbol{\epsilon}^{(D-1)}$ on $\mathcal{N}$ via $\boldsymbol{k} \wedge \boldsymbol{\epsilon}^{(D-1)} = \boldsymbol{\epsilon}^{(D)}$. (Note that since $q_{ij}$ is degenerate, we do not obtain a unique Levi-Civita connection or volume form from the induced metric alone.) We define the second fundamental form, $K^{(k)}_{ij}$, of $\mathcal{N}$ relative to $k^a$ by
\begin{equation}
K^{(k)}_{ij} = \frac{1}{2}\lie_kq_{ij}
\end{equation}
The ``surface gravity'' (or ``non-affinity''), $\kappa^{(k)}$, of $k^a$, which measures the failure of $k^a$ to be affinely parameterized, is given by
	\begin{equation}
	k^b\nabla_bk^a = \kappa^{(k)}k^a
	\label{NonAffinity}
	\end{equation}
 
The tensor field $q_{ij}$ on $\mathcal{N}$ satisfies\footnote{Note that we use the notation $k^i$ when we wish to view $k^a$ as a vector in the tangent space to $\mathcal{N}$ rather than a vector in the tangent space to $\mathcal{M}$.} $q_{ij} k^j = q_{ji}k^j = 0$ and is thus, in this sense, a ``lower dimensional object." It is useful to define a tensor algebra $\mathsf{W}$ of such quantities as follows\footnote{See P.221-222 of \cite{Wald}}: At each point $p \in \mathcal{N}$, let $\hat{\mathcal V}_p$ be the equivalence class of tangent vectors to $\mathcal{N}$ where two vectors are equivalent if they differ by a multiple of $k^i$. The dual space, $\hat{\mathcal V}^*_p$, to $\hat{\mathcal V}_p$ may then be identified with dual vectors $\mu_i$ at $p$ that satisfy $\mu_i k^i = 0$. Let $\mathsf{W}_p$ be the tensor algebra over $\hat{\mathcal V}_p$ and $\hat{\mathcal V}^*_p$ and let $\mathsf{W}$ be the space of such tensor fields on $\mathcal{N}$. We will denote tensor fields in $\mathsf{W}$ with capitalized early Latin indices, i.e., the same notation as we would use for tensor fields on a cross-section, $\mathcal{S}$, of $\mathcal{N}$. It is not unreasonable to do this since the restriction to any cross-section $\mathcal{S}$ of a tensor field in $\mathsf{W}$ is in natural correspondence with a tensor field on $\mathcal{S}$, and it will be convenient to avoid having to make a choice in every instance of whether we are viewing it as a tensor field on $\mathcal{S}$ or as the restriction of a tensor field in $\mathsf{W}$.

The tensor fields $q_{ij}$ and $K^{(k)}_{ij}$ are in $\mathsf{W}$ and will be denoted as $q_{AB}$ and $K^{(k)}_{AB}$ when thought of in this manner. Since $k^a$ is hypersurface orthogonal, the second fundamental form obeys $K^{(k)}_{[AB]} = 0$, and we can decompose it as 
	\begin{equation}
	K^{(k)}_{AB} = \frac{1}{(D-2)}\, \vartheta^{(k)} q_{AB} + \sigma^{(k)}_{AB}
	\label{parallel}
	\end{equation}
	where
	\begin{equation}
	\vartheta^{(k)} = \mathcal{D}_ik^i
	\label{parallelexpansion}
	\end{equation}
	is known as the \emph{expansion} of $\mathcal{N}$, and the symmetric, trace-free\footnote{$\sigma_{AB}$ is trace-free with respect to the unique inverse, $q^{AB}$, of $q_{AB}$ in $\mathsf{W}$.} tensor $\sigma^{(k)}_{AB}$ is known as the \emph{shear} of $\mathcal{N}$. To avoid confusion with the expansion and shear of the families of null geodesics \emph{transverse} to $\mathcal{N}$ that we shall consider later, we will refer to these quantities as the ``parallel expansion" and ``parallel shear" of $\mathcal{N}$ in the following.

It is very useful to introduce {\em Gaussian null coordinates (GNCs)} \cite{GNC1,GNC2,HIW} covering a neighborhood 
of $\mathcal{N}$ in spacetime. To do so, we first fix a choice of normal vector field, $k^a$, on $\mathcal{N}$ and choose a cross-section, $\mathcal{S}$, of $\mathcal{N}$. We then choose coordinates $x^A = (x^1, \dots, x^{D-2})$ on $\mathcal{S}$. We will, of course, in general need more than one coordinate patch to cover $\mathcal{S}$, but, since it is entirely 
straightforward to sew these patches together in the usual way, we will treat the coordinates $x^A$ in our discussion below as though they cover all of $\mathcal{S}$. We extend the coordinates $x^A$ to $\mathcal{N}$ by taking them to be constant along the orbits of $k^i$, and we define the coordinate $v$ on $\mathcal{N}$ by setting $v=0$ on $\mathcal{S}$ and setting $\lie_k v = 1$. Each surface of constant $v$ then defines a cross-section, $\mathcal{S}_v$, of  $\mathcal{N}$. We then define the transverse vector field, $\ell^a$, on  $\mathcal{N}$ by the conditions that $\ell^a$ is a past directed and null, orthogonal to each of the $\mathcal{S}_v$, and normalized via
	\begin{equation}
	k_a\ell^a = 1
	\end{equation}
These conditions uniquely determine $\ell^a$ on $\mathcal{N}$. We then extend $\ell^a$ off $\mathcal{N}$ geodesically ($\ell^b\nabla_b\ell^a = 0$) and let $r$ denote the affine parameter along these geodesics with $r|_{\mathcal{N}} = 0$. Finally, we extend the coordinates $(v, x^A)$ off of $\mathcal{N}$ by holding them constant along orbits of $\ell^a$. The quantities $(v, r, x^A)$ define coordinates in some open neighborhood of ${\mathcal{N}}$. These are the desired Gaussian null coordinates. Note that, given $\mathcal{N}$, the construction of Gaussian null coordinates requires a choice of normalization of $k^a$ and a choice of cross-section $\mathcal{S}$, but it is otherwise unique up to a (for our purposes, irrelevant) choice of coordinates on $\mathcal{S}$.

The metric in Gaussian null coordinates takes the form
	\begin{equation}
	\label{GNF}
	ds^2 = 2\,dvdr -2r\alpha\,dv^2 -2r\beta_A\,dvdx^A + \gamma_{AB}\,dx^Adx^B
	\end{equation}
with $\alpha$, $\beta_A$, and $\gamma_{AB}$ smooth.
Note that we have 
	\begin{equation}
\ell^a = (\partial/\partial r)^a
	\end{equation}
everywhere, and we have, on ${\mathcal{N}}$, that 
	\begin{equation}
k^a \ \widehat{=} \ (\partial/\partial v)^a \, , \qquad \alpha \ \widehat{=} \ \kappa^{(k)}
	\label{alphakappa}
	\end{equation}
where $\,\widehat{=}\,$ denotes that equality holds when both sides are restricted to ${\mathcal{N}}$. The inverse metric takes the form 
\begin{equation}
g^{ab} = 2\,k^{(a}\ell^{\,b)} + r\inp{2\alpha + r\beta_A\beta^A}\ell^a\ell^b + 2r\beta^A\ell^{(a}X^{b)}_A + \gamma^{AB}X_A^aX_B^b
\label{GNCInverse}
\end{equation} 
where $\gamma^{AB}$ is the inverse of $\gamma_{AB}$, $\beta^A \coloneqq \gamma^{AB}\beta_B$ and $X^a_A \coloneqq (\partial/\partial x^A)^a$.

We have previously defined the expansion and shear of the null hypersurface $\mathcal{N}$ in \cref{parallel}. For any cross-section $\mathcal{S}$, the null geodesics orthogonal to $\mathcal{S}$ generated by $\ell^a$ also comprise a null hypersurface. We call the expansion, $\vartheta^{(\ell)}$, and shear, $\sigma^{(\ell)}_{ab}$, of this hypersurface the \emph{transverse expansion} and \emph{transverse shear} respectively. In Gaussian null coordinates, we have that, on $\mathcal{N}$,
	\begin{equation}
	\vartheta^{(k)} \ \widehat{=} \ \frac{1}{2}\gamma^{AB}\partial_v \gamma_{AB} = \pd{\ln(\sqrt{\gamma})}{v} \quad\text{and}\quad \vartheta^{(\ell)} \ \widehat{=} \ \frac{1}{2}\gamma^{AB}\partial_r\gamma_{AB} = \pd{\ln(\sqrt{\gamma})}{r}
	\label{Expansions}
	\end{equation}
	
	\subsection{Symmetries of a Null Hypersurface}
	\label{SymmetriesNull}

 Chandrasekaran, Flanagan, and Prabhu (CFP) \cite{CFP} have defined the notion of the symmetry group, $\mathscr{G}_{\mathcal{N}}$, of a null surface, $\mathcal{N}$, as the group of diffeomorphisms that preserve the ``universal intrinsic structure'' of $\mathcal{N}$, defined as the equivalence class of $(k^i,\, \kappa^{(k)})$, with the equivalence relation given by 
		\begin{equation}
		(k^i,\,\kappa^{(k)}) \sim \inp{e^fk^i,\, e^f(\kappa^{(k)} + \lie_kf)}, \qquad f \in C^{\infty}(\mathcal{N})
		\end{equation}
This symmetry group has the structure				
	\begin{equation}
	\mathscr{G}_{\mathcal{N}} \simeq \mathrm{Diff}(\bar{\mathcal{S}})\ltimes \mathscr{S}
	\label{symgroup} 
	\end{equation} 
where $\bar{\mathcal{S}}$ is the manifold of orbits of $k^a$. Our interest here is in the subgroup, $\mathscr{S}$, of ``generalized supertranslations'' of $\mathcal{N}$. If $V$ denotes the coordinate $v$ of the previous subsection with $k^a$ chosen to be affinely parameterized, then $\mathscr{S}$ consists of all diffeomorphisms of $\mathcal{N}$ of the form
	\begin{equation}
	\ (V,x^A) \mapsto \big(e^{\phi_1(x^A)}V + \phi_2(x^A),\, x^A\big),
	\label{cfpsuper}
	\end{equation} 
where $\phi_1$ and $\phi_2$ are arbitrary smooth functions on $\bar{\mathcal{S}}$. The generalized supertranslations are 
infinitesimally generated by the vector fields
	\begin{equation}
	\xi^i = (\tilde{\phi}_2 + V\phi_1)\inp{\pd{}{V}}^i \quad\text{where}\quad
	\tilde{\phi}_2 = \inp{\frac{\phi_1}{e^{\phi_1}-1}}\phi_2
	\label{infsuper}
	\end{equation}
Equivalently, we may write
	\begin{equation}
	\xi^i = f k^i 
	\label{gensuper}
	\end{equation} 
where
	\begin{equation}
	\lie_k(\lie_k + \kappa^{(k)})f = 0
	\end{equation} 
We refer to generalized supertranslations with $\phi_1 = 0$ as \emph{affine supertranslations} and refer to the supertranslations with $\phi_2 = 0$ as \emph{Killing supertranslations}\footnote{The reason for this terminology will be made clear at the beginning of the next section.}. For the affine supertranslations, we have $(\lie_k + \kappa^{(k)})f = 0$.

CFP have shown \cite{CFP} that in general relativity, charges and fluxes can be associated with symmetries of $\mathcal{N}$ by the same procedure as used in \cite{WZ} to associate charges and fluxes with symmetries of $\mathcal{I}^+$. Our interest here is in the charges and fluxes associated with a generalized supertranslation, $\xi^i$, of \cref{gensuper}. The local supertranslation charge, $\mathcal{Q}_{(\xi)}^{\mathrm{loc}}$, on a cross-section $\mathcal S$ is given by \cite{CFP}
	\begin{equation}
	\mathcal{Q}_{(\xi)}^{\mathrm{loc}}(\mathcal{S}) = \frac{1}{8\pi}\int_{\mathcal{S}}\insb{(\vartheta^{(k)}-\lie_k-\kappa^{(k)})f}\boldsymbol{\epsilon}^{(D-2)}
	\label{CFPcharge}
	\end{equation}
where $\boldsymbol{\epsilon}^{(D-2)}$ is the pullback of the spacetime volume form to $\mathcal{S}$. If we let $\Delta\mathcal{N}$ denote the region of $\mathcal{N}$ bounded by the cross-sections $\mathcal{S}_0$ and $\mathcal{S}_1$, the integrated supertranslation flux, $\mathcal{F}_{(\xi)}(\Delta\mathcal{N})$, through this region is given by \cite{CFP}
	\begin{equation}
	\mathcal{F}_{(\xi)}(\Delta\mathcal{N}) = \frac{1}{8\pi}\int_{\Delta\mathcal{N}}\inp{q^{AC}q^{BD}\sigma^{(k)}_{AB}\,\sigma^{(k)}_{CD}-\frac{1}{2}(\vartheta^{(k)})^2}f\,\boldsymbol{\epsilon}^{(D-1)}
	\label{HSTFlux}
	\end{equation}
Note that for a given supertranslation $\xi^i$,  
this formula does not depend upon the choice of normal $k^i$. Namely, if $k^i \to e^h k^i$ with $h$ an arbitrary smooth
function on $\mathcal{N}$, then ${\sigma^{(k)}_{AB} \to e^h \sigma^{(k)}_{AB}}$, $\vartheta^{(k)} \to e^h\vartheta^{(k)}$, $f \to e^{-h} f$, and $\boldsymbol{\epsilon}^{(D-1)} \to e^{-h}\,\boldsymbol{\epsilon}^{(D-1)}$, so $\mathcal{F}_{(\xi)}(\Delta\mathcal{N})$ remains invariant.

	\section{Killing Horizon Foliations}
	\label{KHFol}

A Killing horizon $(\mathcal{N}, \chi^a)$ is a null surface $\mathcal{N}$ with a normal 
that is the restriction of a spacetime Killing vector field, $\chi^a$, to $\mathcal{N}$. If Einstein's equation holds with matter satisfying the dominant energy condition, then the surface gravity, $\kappa$, of $\mathcal{N}$ with respect to the normal $\chi^a$ (see \cref{NonAffinity}) must be constant on $\mathcal{N}$ \cite{BHThermo}. If $\kappa$ is constant and $\kappa \neq 0$, then the Killing horizon is said to be \emph{non-extremal}\footnote{Any such non-extremal Killing horizon can always be extended (if necessary) to a bifurcate Killing horizon \cite{RaczWald} (see \cref{tphiReflectionFoliation} below).}.

As previously noted, in Gaussian null coordinates on any null surface $\mathcal{N}$ constructed with respect to some normal $k^a$, we have that $k^a = (\partial/\partial v)^a$ on $\mathcal{N}$. In the case of a Killing horizon, if we choose Gaussian null coordinates with $k^a = \chi^a|_{\mathcal{N}}$ then the construction of the coordinates $r$ and $x^A$ on spacetime will be invariant under the action of $\chi^a$. It follows that $\chi^a = (\partial/\partial v)^a$ throughout the domain where the Gaussian null coordinates are defined.
In particular, the Gaussian null coordinate $v$ will be a Killing parameter on spacetime, i.e., $\chi^a \nabla_a v = 1$. Furthermore, on $\mathcal{N}$, the quantity $V = \exp (\kappa v)$ will be an affine parameter along the null generators of $\mathcal{N}$. The affine supertranslations of \cref{SymmetriesNull} are then given by
		\begin{equation}
		V \mapsto V + \phi_2(x^A)
		\end{equation} 
whereas the Killing supertranslations are given by
		\begin{equation}
		v \mapsto v + \kappa^{-1}\phi_1(x^A)
		\label{KillingST}
		\end{equation} 

A theorem of Hawking \cite{HE, HIW} asserts that the event horizon of a stationary black hole must be a Killing horizon. For the analysis of black hole memory, we are interested in a situation where a black hole initially is approximately\footnote{Unless the black hole is exactly stationary at all times, Raychaudhuri's equation implies that the expansion must be positive at early times, so the black hole cannot be exactly stationary at early times.} stationary, goes through a dynamical era, and then becomes (asymptotically) stationary again at late times. Thus, in the early and late time eras, the event horizon of the black hole should be well described by a Killing horizon. As discussed in \cref{memnullinf}, memory at $\mathcal{I}^+$ can be characterized by the supertranslation that relates the foliations of ``good cuts'' of $\mathcal{I}^+$ in the early and late time stationary eras. Thus, a similar notion of black hole memory could be defined if we can define a similar foliation of ``good cuts'' of a Killing horizon and show that they are related by a supertranslation. The good cuts of $\mathcal{I}^+$ are characterized by vanishing leading order transverse shear. However, a general Killing horizon does not, in general, admit any cross-sections of vanishing transverse shear\footnote{This can be seen from \eqref{ShearEq} below, which yields an overdetermined equation for the function $f$ if one sets the left hand side to zero.}. Therefore, we seek an alternative criterion for defining preferred foliations of a Killing horizon. 

In this section, we will restrict consideration to non-extremal Killing horizons with compact cross-sections. 
In the following subsections, we will analyze three prescriptions for defining preferred foliations of
such non-extremal Killing horizons. In \cref{ConstantExpansionFoliation}, we show that the requirement that the cross-sections have uniform transverse expansion gives rise to a unique foliation. In \cref{tphiReflectionFoliation}, we consider Killing horizons in spacetimes admitting a $t$--$\phi$ reflection symmetry and use that symmetry to define a 
unique foliation. In \cref{ABLFoliation}, we consider a prescription previously given by \cite{ABL1, ABL2} which requires the one-form $\beta_A\,\mathrm{d}x^A$ in \cref{GNF} to be divergence free on $\mathcal{N}$. We show that the prescriptions of \cref{tphiReflectionFoliation} (when defined) and \cref{ABLFoliation} are equivalent, but that they differ, in general, from the prescription of \cref{ConstantExpansionFoliation}. Thus, we have two distinct candidates for the  
``good cuts" of a Killing horizon to use 
in defining a notion of black hole memory.

It is worth noting that none of the prescriptions below can be applied to get preferred cross-sections of $\mathcal{I}^+$ in stationary eras. The transverse expansion of {\em any} cross-section of $\mathcal{I}^+$ is given by $(D-2)/r$ at order $1/r$, whereas the transverse expansion is entirely gauge dependent at order $1/r^2$. Thus, the condition of uniform transverse expansion cannot be used to pick out preferred cross-sections of $\mathcal{I}^+$. The construction of the $t$--$\phi$ reflection foliation depends on having a regular bifurcation surface and the construction of the ABL foliation is based on properties of $\beta_A$, which vanishes at $\mathcal{I}^+$. Thus, neither the $t$--$\phi$ construction nor the ABL construction can be applied to $\mathcal{I}^+$ (nor can they be applied to extremal Killing horizons). Thus, none of the constructions below can be used to replace the vanishing of the transverse shear as a condition defining ``good cuts'' of $\mathcal{I}^+$.
	
\subsection{The Uniform Expansion Foliation}
\label{ConstantExpansionFoliation} 

Let $(\mathcal{N}, \chi^a)$ be a non-extremal Killing horizon of constant surface gravity $\kappa$, and let $(v,r,x^A)$ be Gaussian null coordinates constructed with respect to the normal $k^a \ \widehat{=} \ \chi^a$ and some arbitrary cross-section, $\mathcal{S}$, of $\mathcal{N}$. In these coordinates, the transverse expansion of $\mathcal{S}$ is given by
\begin{equation}
	\vartheta^{(\ell)} = \frac{1}{2}\,\gamma^{AB}\partial_r\gamma_{AB}\Big|_{r = 0}
	\label{expansioneqs}
\end{equation}
(see \cref{Expansions}). Note that any other cross-section, $\tilde{\mathcal{S}}$, of $\mathcal{N}$ can be written in these coordinates as the surface on which $\tilde{v} = v + f = 0$, where $f$ is some smooth function on $\mathcal{N}$ with $\lie_{\chi}f = 0$. In other words, any other cross-section $\tilde{\mathcal{S}}$ can be written as the image of $\mathcal{S}$ under a Killing supertranslation. We wish to find such an $\tilde{\mathcal{S}}$ whose transverse expansion, $\tilde{\vartheta}^{(\tilde{\ell})}$, is uniform over all of $\tilde{\mathcal{S}}$. We can then extend $\tilde{\mathcal{S}}$ to a $\chi^a$-invariant foliation of $\mathcal{N}$ via rigid transport along the Killing isometry. 

To this end, let $(\tilde{v}, \tilde{r}, \tilde{x}^A)$ be new Gaussian null coordinates constructed with respect to $k^a \ \widehat{=} \ \chi^a$ and the new cross-section $\tilde{\mathcal{S}}$. We have $\tilde{v} \ \widehat{=} \ v + f$ for some $f$ with $\lie_{\chi}f = 0$ and we have $\tilde{r} \ \widehat{=} \ 0$, where, again, $\widehat{=}\,$ indicates
equality on $\mathcal{N}$. We choose $\tilde{x}^A \ \widehat{=} \ x^A$. In order to calculate the transverse expansion $\tilde{\vartheta}^{(\tilde{\ell})}$ of $\tilde{\mathcal{S}}$ using the ``tilded" analog of \cref{expansioneqs}, we need to relate the new coordinate functions, $(\tilde{v}, \tilde{r}, \tilde{x}^A)$, to the old coordinate functions, $(v,r,x^A)$, to leading order in $r$. We write
	\begin{align}
	\tilde{v} &= v + f + rF^{(v)} + O(r^2) \label{Fv}\\
	 \tilde{r} &= rF^{(r)}  + O(r^2) \label{Fr}\\
	\tilde{x}^A &= x^A + rF^{(A)} + O(r^2) \label{FA}
	\end{align}
where the expansion coefficients $F^{(\mu)}$ are, of course, independent of $r$. Since $\chi^a = (\partial/\partial v)^a = (\partial/\partial \tilde{v})^a$ (as both the old and new coordinates are Gaussian null coordinates with $k^a = \chi^a$), we also have $\lie_\chi F^{(\mu)} = 0$, i.e., the coefficients $F^{(\mu)}$ are functions of $x^A$ only.

It is convenient to work with the inverse metric $g^{ab}$ rather than $g_{ab}$. In the original Gaussian null coordinates, $g^{ab}$ takes the form \cref{GNCInverse}. Applying the coordinate transformation \cref{Fv}-\cref{FA} and imposing the conditions
	\begin{equation}
	\tilde{g}^{\tilde{v}\tilde{r}} = 1, \qquad \tilde{g}^{\tilde{v}\tilde{v}} = 0, \qquad \tilde{g}^{\tilde{v}\tilde{A}} = 0	
	\label{nonperturbativeconstraints}
	\end{equation}
we obtain\footnote{Here we view $f$ as a function on $\mathcal{S}$, with $D_A$ the derivative operator on $\mathcal{S}$ associated with the pullback metric $\gamma_{AB}$, and we raise and lower capital Latin indices with $\gamma^{AB}$ and $\gamma_{AB}$.}
	\begin{equation} 
	F^{(v)} = -\frac{1}{2}\,D^AfD_Af, \qquad  F^{(r)} = 1, \qquad F^A = -D^Af
	\label{Fs}
	\end{equation}
The remaining components of the inverse metric in the new Gaussian null coordinates are
\begin{equation}
	\tilde{g}^{\tilde{r}\tilde{r}} = \tilde{r}\inp{2\tilde{\alpha} + \tilde{r}\,\tilde{\gamma}^{\tilde{A}\tilde{B}}\tilde{\beta}_{\tilde{A}}\tilde{\beta}_{\tilde{B}}}, \qquad \tilde{g}^{\tilde{r}\tilde{B}} = 2\tilde{r}\,\tilde{\gamma}^{\tilde{A}\tilde{B}}\tilde{\beta}_{\tilde{B}}, \qquad \tilde{g}^{\tilde{A}\tilde{B}} = \tilde{\gamma}^{\tilde{A}\tilde{B}} = [\tilde{\gamma}_{\,\cdot\,\cdot}^{-1}]^{\tilde{A}\tilde{B}}
	\label{conditions}
\end{equation}
We now apply the coordinate transformation eqs.~(\ref{Fv})-(\ref{FA}), with $F^{(\mu)}$ given by \cref{Fs}, to the components of the inverse metric in the original Gaussian null coordinates and match the result with \cref{conditions}. We obtain
\begin{equation}
	\tilde{\alpha} \ \widehat{=} \  \alpha, \qquad \tilde{\beta}_A \ \widehat{=} \  \beta_A - 2\kappa\,D_Af, \qquad \tilde{\gamma}_{AB} \ \widehat{=} \  \gamma_{AB}
	\label{orderzerochanges}
\end{equation}	
and (by a lengthier calculation)
	\begin{equation}
	\partial_{\tilde{r}}\tilde{\gamma}_{AB} \ \widehat{=} \  \partial_r\gamma_{AB} + 2\,D_AD_Bf -2\kappa\,D_AfD_Bf + 2\,\beta_{(A}D_{B)}f
	\label{ShearEq}
	\end{equation}	
Using \cref{expansioneqs} and \eqref{ShearEq}, we see that the expansion of $\tilde{\mathcal{S}}$ is given by
	\begin{equation}
	\tilde{\vartheta}^{(\tilde{\ell})}\big|_{\tilde{\mathcal{S}}}
	= \insb{\vartheta^{(\ell)} + (D^AD_Af -\kappa D^AfD_Af + \beta^AD_Af)}\Big|_{\mathcal{S}} \vspace{1mm}
	\label{NewExpansion}
	\end{equation}
Thus, $\mathcal{N}$ will possess a cross-section of uniform transverse expansion if and only if one can find a function, $f$, on $\mathcal{S}$ such that 
	\begin{equation}
	-D^AD_Af+\kappa\,D^Af\,D_Af - \beta^AD_Af = (\vartheta^{(\ell)}-\tilde{\vartheta})
	\label{NonlinearEquation}
	\end{equation}
with $\tilde{\vartheta}$ constant (i.e., independent of $x^A$). \Cref{NonlinearEquation} is nonlinear, but the change of variables
	\begin{equation}
	f \to -\kappa^{-1}\ln(F)
	\end{equation}
converts it to the linear equation
	\begin{equation}
	\mathbf{L}F = \kappa\tilde{\vartheta}\,F, \qquad \mathbf{L} = \inp{-D^AD_A-\beta^AD_A+\kappa\,\vartheta^{(\ell)}}
	\label{evL}
	\end{equation}
subject to the restriction that $F$ be real and strictly positive. Since we allow $\tilde{\vartheta}$ to be any real constant, \cref{evL} will be solved if and only if we can find an eigenfunction, $F$, of $\mathbf{L}$ that is everywhere real and positive. Since $\mathbf{L}$ is a second-order linear elliptic operator on a compact Riemannian manifold $(\mathcal{S}, \gamma_{AB})$, it is well known \cite{Mars} that it has a unique minimal real eigenvalue $\lambda$---the so called \emph{principal eigenvalue}---with real, positive eigenfunction $F$. Thus, \cref{evL} can be solved. The desired foliation of $\mathcal{N}$ by cross-sections of uniform transverse expansion will then be given by the surfaces of constant $\tilde{v} = v-\kappa^{-1}\ln(F)$.

It is not difficult to show that this foliation is unique. To prove this, let $\tilde{\mathcal{S}}$ be a cross-section of $\mathcal{N}$ which generates a foliation of constant transverse expansion $\tilde{\vartheta}$, and let $\mathcal{S}'$ be any other cross section of $\mathcal{N}$ that also has constant transverse expansion. Expressing $\mathcal{S}'$ as the surface $\tilde{v} = v-\kappa^{-1}\ln(\tilde{F})$ for some strictly positive $\tilde{F}$, we find, by the same steps as led to \cref{evL}, that $\tilde{F}$ must satisfy
	\begin{equation}
	\tilde{\mathbf{L}}\tilde{F} = \kappa(\vartheta'-\tilde{\vartheta})\tilde{F}, \qquad\tilde{\mathbf{L}} = \inp{-D^AD_A-\tilde{\beta}^AD_A}
	\label{uniqueeq}
	\end{equation}
where now both $\tilde{\vartheta}$ and $\vartheta'$ are constant. Since $\tilde{\mathcal{S}}$ is compact, $\tilde{F}$ attains maximum and minimum values on $\tilde{\mathcal{S}}$. By \cref{uniqueeq}, at any critical point $p^*$, $\tilde{F}$ satisfies 
	\begin{equation}
	-D^AD_A\tilde{F}\big|_{p^*} = \kappa(\vartheta'-\tilde{\vartheta})\tilde{F}(p^*)
	\label{cp}
	\end{equation}  
If $\vartheta' >  \tilde{\vartheta}$, then the right side of \cref{cp} is everywhere positive and we obtain a contradiction when $p^*$ is chosen to be a point at which $\tilde{F}$ achieves its minimum value. If $\vartheta' <  \tilde{\vartheta}$, then the right side of \cref{cp} is everywhere negative and we obtain a contradiction when $p^*$ is chosen to be a point at which $\tilde{F}$ achieves its maximum value. Thus, we must have $\vartheta' =  \tilde{\vartheta}$, in which case \cref{uniqueeq} reduces to 
	\begin{equation}
	\tilde{\mathbf{L}}\tilde{F} = 0
	\label{evL0}
	\end{equation}
Since $\tilde{F}$ attains its global maximum on $\tilde{\mathcal{S}} = \rm{int}(\tilde{\mathcal{S}})$, it follows immediately from the strong maximum principle that the only solutions to \cref{evL0} are $\tilde{F} = \rm{const}$. Thus, $\mathcal{S}'$ must be a member of the original transverse expansion foliation, and this foliation is unique.

\subsection{The $t$--$\phi$ Reflection Foliation}
\label{tphiReflectionFoliation}
As previously mentioned, the event horizon of a stationary black hole must be a Killing horizon \cite{HE, HIW}. If the black hole is rotating, then the horizon Killing field cannot coincide with stationary Killing field, and the spacetime must be axisymmetric as well as stationary \cite{HIW}. In four spacetime dimensions, one can prove \cite{r1,r2} that any stationary and axisymmetric (asymptotically flat) solution of Einstein's equation in vacuum will carry a discrete ``$t$--$\phi$ reflection isometry," which simultaneously reverses the stationary and axisymmetric Killing fields. This proof does not generalize to higher dimensions, but under additional assumptions, the existence of a $t$--$\phi$ reflection isometry can be proven to hold in arbitrary spacetime dimension \cite{highertphi}. Thus, it is of interest to consider Killing horizons 
embedded within spacetimes admitting a reflection isometry of this type.

Let $(\mathcal{N}, \chi^a)$ be any (connected) nonextremal Killing horizon with constant surface gravity $\kappa$. Following \cite{RaczWald}, we can define a Kruskal-type extension of a neighborhood of $\mathcal{N}$ to obtain a spacetime with a bifurcate Killing horizon. To do so, we start with Gaussian null coordinates \cref{GNF} with $k^a \ \widehat{=} \ \chi^a$. In these coordinates, all metric components are independent of $v$. We define\footnote{Note that $U$ and $V$ are reversed relative to the conventions of \cite{RaczWald}.}
	\begin{equation}
 V = e^{\kappa v} \, ,  \qquad U = -re^{-\kappa v}\exp\insb{\kappa\intf{0}{r}\frac{1}{r'}\inp{\frac{1}{\alpha(r', x^A)}-\frac{1}{\kappa}}\mathrm{d}r'}
 	\label{UVdef}
	\end{equation}
and eliminate the coordinates $(v,r)$ in favor of $(V,U)$. Note that the integrand in the expression for $U$ is smooth at $r' = 0$ by virtue of \cref{alphakappa}, so $U$ is well defined and smooth. The metric then takes the form
	\begin{equation}
	ds^2 = 2\,G\,dUdV + 2\,UH_A\,dVdx^A + \gamma_{AB}\,dx^Adx^B
	\label{KC}
	\end{equation}
where all metric components depend on $U$ and $V$ only through the combination $UV$. Note that the original coordinate $r$ can be expressed as a smooth function of $UV$ and $x^A$, ${r=r(UV,x^A)}$.

The range of the coordinate $V$ of \cref{UVdef} is $V>0$. We can extend the spacetime by allowing $-\infty < V < \infty$. The resulting spacetime contains a ``bifurcate Killing horizon'' as the null surfaces $V=0$ and $U=0$ are both Killing horizons. The $(D-2)$-dimensional intersection surface, $\mathcal{B}$, at $U=V=0$ is called the \emph{bifurcation surface}.

We now assume that $(\mathcal{N}, \chi^a)$ is the event horizon of a black hole, and that the exterior of the black hole is stationary with Killing field $\xi^a$ and is axisymmetric with axial Killing fields $\psi_{\Lambda}^a$, $\Lambda = 1, \dots, p$, where $p$ is the number of axial Killing fields. We choose the cross-section $\mathcal S$ of the Gaussian null coordinates \cref{GNF} to be invariant under the axial isometries and choose the coordinates $x^A$ to be of the form $x^A = (\vphi^{\Lambda}, \theta^{\alpha})$, with
	\begin{equation}
	\lie_{\psi_{\Lambda}}\vphi^{\Gamma} = \delta^{\Lambda\Gamma} \, , \qquad \lie_{\psi_{\Lambda}}\theta^{\alpha} = 0
	\end{equation} 
(but with $\theta^{\alpha}$ otherwise arbitrary). All quantities appearing in the metric forms \cref{GNF} and \cref{KC} will then be independent of $\vphi^{\Lambda}$.

It is useful to define a metric $\Phi_{\Lambda \Sigma}$ (which depends on the spacetime point) on the vector space of axial Killing fields via
	\begin{equation}
	\Phi_{\Lambda \Sigma} = g_{ab} \psi_{\Lambda}^a \psi_{\Sigma}^b
	\end{equation} 
We may then define
	\begin{equation}
	\Psi^{ab} = \Phi^{\Lambda \Sigma} \psi_{\Lambda}^a \psi_{\Sigma}^b
	\label{Psiab}
	\end{equation} 
where $\Phi^{\Lambda \Sigma}$ is the inverse of $\Phi_{\Lambda \Sigma}$. Since $\psi_{\Lambda}^a$ is tangent to the surfaces of constant $(v,r)$ in the original Gaussian null coordinates \cref{GNF}, we may view \eqref{Psiab} as a tensor field $\Psi^{AB}$ on these surfaces. We define
	\begin{equation}
	A(r, \theta^\alpha) = \frac{r}{2 \alpha}\,\Psi^{AB} \beta_A \beta_B
	\label{Adef}
	\end{equation} 

We now further assume that the spacetime possesses a $t$--$\phi$ reflection isometry. The reflection isometry implies \cite{RaczWald, highertphi} that the exterior region can be foliated by spacelike hypersurfaces, $\{\Sigma_t\}$, given by the level sets of a function $t$ with the properties that $\lie_{\xi}t = 1$, $\lie_{\psi_{\Lambda}}t = 0$, and that $\nabla^a t$ lies in the span of $\xi^a$ and the $\psi_{\Lambda}^a$.

Paralleling the derivation given in \cite{RaczWald} for the $4$ spacetime dimensional case, we find that the hypersurfaces $\{\Sigma_t\}$ are given by the solution set of the equation
	\begin{equation}
	V + U\,e^{2\kappa\inp{t-\tilde{H}(UV, \theta^{\alpha})}} = 0
	\label{UVeq}
	\end{equation}
where $\tilde{H}$ is given by\footnote{Note that there is a typo in the corresponding equation, (57), of \cite{RaczWald}, where the factor of $1/(2r'\alpha)$ in the integrand was omitted.}
	\begin{equation}
	\tilde{H}(UV, \theta^{\alpha}) = \intf{0}{r(UV, x^A)}\frac{1}{2r'\alpha(r', \theta^{\alpha}) }\cdot\frac{A(r', \theta^{\alpha})}{1 + A(r', \theta^{\alpha})}\,\mathrm{d}r' + B(\theta^{\alpha})
	\end{equation}
where $A$ was defined by \cref{Adef} and $B(\theta^{\alpha})$ is determined up to addition of a constant by\footnote{That a $B$ satisfying \cref{betatphi} exists is a consequence of our initial {\em assumption} of a $t$--$\phi$ reflection isometry, which implied the existence of the foliation $\{\Sigma_t\}$.}
	\begin{equation}
	D_A B = \frac{1}{2\kappa}\,\Theta_{AB}\,\beta^{B}\Big|_{\mathcal{H}^+}
	\label{betatphi}
	\end{equation}
where $\Theta_{AB}$ is the tensor field on ${\mathcal{H}^+}$ defined by $\Theta_{AB} = \gamma_{AB} - \Psi_{AB}$. 
	
The key point is that $\tilde{H}$ is a smooth function of $(UV, \theta^\alpha)$. It follows \cite{RaczWald} that, for each $t$, the solution to \cref{UVeq} extends smoothly through $V=U=0$ and, thus, that each $\Sigma_t$ extends smoothly through the bifurcation surface $\mathcal{B}$. The normal, $n_a$, to $\Sigma_t$ at $\mathcal{B}$ is given by
	\begin{equation}
	n_a\big|_{\mathcal{B}}\, = -\frac{1}{\sqrt{2\kappa}}\inp{e^{-\kappa(t-B(\theta^{\alpha}))}(\mathrm{d}V)_a + e^{\kappa(t-B(\theta^{\alpha}))}(\mathrm{d}U)_a}
	\end{equation}
We can define a unique, future-directed normal $\eta^a$ to the (extended) Killing horizon at $\mathcal{B}$ by the condition
	\begin{equation}
	\eta^a n_a\big|_{\mathcal{B}} = -\sqrt{2\kappa}
	\label{normconv}
	\end{equation}
so that, in our coordinates, we have at $\mathcal{B}$
	\begin{equation}
	\eta^a\big|_{\mathcal{B}} = e^{2\kappa\inp{t-B(\theta^{\alpha})}}\inp{\pd{}{V}}^a
	\label{nB}
	\end{equation}
We then may define a foliation of $\mathcal{N}$ by the surfaces of constant affine parameter $\lambda$ along the null geodesics determined by $\eta^a$, with $\lambda = 0$ at $\mathcal{B}$. This foliation is independent of the choice of $t$--$\phi$ reflection invariant surface $\Sigma_t$, since a contant shift of $t$ merely rescales $\eta^a$ by a constant and so yields the same foliation. Thus, we obtain the desired unique foliation defined by the $t$--$\phi$ reflection symmetry.
	
	\subsection{The ABL Foliation} 
	\label{ABLFoliation}

A third prescription for foliating a nonextremal Killing horizon (which we will call the ``ABL foliation") was given by \cite{ABL1,ABL2}.
The defining condition of the ABL foliation is that a cross-section, $\mathcal{S}$, be chosen so that, in Gaussian null coordinates constructed with respect to the horizon Killing field and $\mathcal{S}$, the one-form $\beta_A$ of \cref{GNF} satisfies
	\begin{equation}
	D^A\beta_A \ \widehat{=} \, \ 0
	\label{divbeta}
	\end{equation}
The ABL foliation is then the foliation, $\mathcal{S}_v$, of these Gaussian null coordinates.

As discussed in \cref{ConstantExpansionFoliation}, a change in the cross section 
${\mathcal S}$ 
used in the GNC construction corresponds to $v \mapsto v + f$ on $\mathcal N$ with $\lie_kf = 0$. Under the corresponding change in GNC, $\beta_A$ transforms as $\beta_A \mapsto \beta_A-2\kappa\,D_A f$ (see \cref{orderzerochanges} above). Thus, to implement this prescription, we must solve the Poisson equation
	\begin{equation}
	D^AD_Af = \frac{1}{2\kappa}\,D^A\beta_A 
	\end{equation}
for $f$ on ${\mathcal S}$. Since $\int_{\mathcal S} D^A\beta_A = 0$, this equation can always be solved \cite{Donaldson} with solution, $f$, unique up to the addition of a constant. Thus, there exists a unique foliation of the Killing horizon determined by \cref{divbeta}.

	\subsection{Comparison of the Foliation Prescriptions}
	\label{FoliationComparison}

In this subsection, we compare the foliation prescriptions of the previous three subsections. We will show that, where defined, the $t$--$\phi$ reflection foliation coincides with the ABL foliation. However, we find that the uniform expansion foliation differs, in general, from the ABL foliation.
	
	\subsubsection{Comparing the $t$--$\phi$ and ABL Foliations} 

Suppose the spacetime possesses a $t$--$\phi$ reflection isometry, so that the $t$--$\phi$ reflection foliation exists. Start with Gaussian null coordinates with respect to a cross-section in this foliation. The construction of the $t$--$\phi$ reflection foliation in these coordinates must then simply give back the original foliation. This implies that, in these coordinates, the null normal \cref{nB} at $\mathcal{B}$ that defines the $t$--$\phi$ reflection foliation must be of the form
	\begin{equation}
	\eta^a\big|_{\mathcal{B}} = c\inp{\pd{}{V}}^a
	\end{equation}
for some constant $c$. Hence, by \cref{nB}, we must have $B(\theta^\alpha) = {\rm constant}$ in these coordinates. Then, by \cref{betatphi}, we have $\Theta_{AB}\beta^{B} \ \widehat{=} \ 0$, which implies that, on $\mathcal{H}^+$, $\beta^A$ must lie in the span of the axial Killing fields, $\psi^A_{\Lambda}$,
	\begin{equation}
	\beta^A \ \widehat{=} \ \sum_\Lambda f^\Lambda(\theta^\alpha) \,\psi^A_{\Lambda}
	\end{equation}
It follows immediately that $D_A \beta^A \ \widehat{=} \ 0$. Thus, when defined, the $t$--$\phi$ reflection foliation coincides with the ABL foliation.
	
	\subsubsection{Comparing the ABL and Uniform Expansion Foliations}
	
In any Gaussian null coordinates on the Killing horizon  $(\mathcal{N}, \chi^a)$ with $k^a \ \widehat{=} \ \chi^a$, the ``angle-angle'' components of the vacuum Einstein's equation read \cite{HIW}
	\begin{equation}
	0 \ \widehat{=} \   R_{AB} \ \widehat{=} \,  -\kappa \,\partial_r\gamma_{AB} + \mathcal{R}_{AB}-D_{(A\,}\beta_{B)}-\frac{1}{2}\,\beta_A\beta_B
	\end{equation} 
Here $R_{AB}$ denotes the $(\partial/\partial x^A)$-components of the spacetime Ricci tensor, while $\mathcal{R}_{AB}$ denotes the Ricci tensor of $\gamma_{AB}$. Taking the trace (with respect to $\gamma^{AB}$), we find that
	\begin{equation}
	0 \ \widehat{=} \  -2\kappa\,\vartheta^{(\ell)} + \mathcal{R} - D^{A\,}\beta_A  -\frac{1}{2}\,\beta^A\beta_A
	\end{equation}
For the ABL foliation, we have $D^{A\,}\beta_A \ \widehat{=} \ 0$. Thus, for the ABL foliation, the transverse expansion is given by
	\begin{equation}
	\vartheta^{(\ell)} \ \widehat{=} \ \frac{1}{2\kappa}\,\Big( \mathcal{R}-\frac{1}{2}\,\beta^A\beta_A \Big)
	\label{ceash}
	\end{equation}
The right hand side of \cref{ceash} is constant over the ABL cross-sections of Schwarzschild, where both the ABL and uniform expansion cross-sections coincide with orbits of spherical symmetry. However, it can be checked that the right hand side is not constant over the ABL cross-sections of Kerr. Thus, the uniform expansion and ABL cross-sections do not agree in general, even when the vacuum Einstein equation is imposed.

Indeed, more generally, $\gamma_{AB}$ and $\beta_A$ are free initial data on a bifurcate Killing horizon for a solution to the vacuum Einstein equation (see theorem 2 of \cite{CRK}). Thus, one can easily construct examples of Killing horizons for which $D^A\beta_A \ \widehat{=} \ 0$ but the right side of \cref{ceash} is not constant (e.g., one could choose $\beta_A=0$ and choose $\gamma_{AB}$ so that $\mathcal{R}$ is not constant). Thus, it is easy to construct examples where the uniform expansion and ABL foliations differ.
	
	\section{Black Hole Memory}
	\label{BHMem}

In this section, we will consider two notions of black hole memory analogous to the notions of memory at null infinity based on geodesic deviation and on supertranslations relating good cuts. We consider a black hole that is initially approximately stationary, goes through a dynamical era, and becomes approximately stationary again at late times. We will assume that the black hole event horizon, $\mathcal{H}^+$, can be well approximated by a nonextremal Killing horizon in the early and late time approximate stationary eras.

In general, during the dynamical era, new horizon generators may be created on $\mathcal{H}^+$. If so, this would cause difficulties for any definition of memory, since some of the late time generators would have no correspondence with the early time era. We will ignore any such issues here and treat $\mathcal{H}^+$ as though it were a smooth null surface. 

\subsection{The Memory Tensor for Black Hole Horizons}
\label{bhmemten}

As discussed in \cref{memnullinf} above, the memory effect at $\mathcal{I}^+$ is defined by considering the relative displacement of test masses (timelike geodesics) near $\mathcal{I}^+$ that are initially stationary, i.e., whose worldlines are initially tangent to $(\partial/\partial u)^a$ in the coordinates of \cref{scricoord}. Our first task in attempting to formulate a notion of black hole memory is to obtain a corresponding family of worldlines whose relative displacements could be used to define memory. Initially stationary timelike geodesics just outside of a stationary black hole will fall into the black hole almost immediately and thus cannot be used to define a notion of memory. Accelerating worldlines outside of a black hole can avoid falling into the black hole, but the relative displacement of these worldlines depends on the choice of acceleration, and there does not appear to be any natural choice during the nonstationary era. However, the null geodesic generators of the future event horizon $\mathcal{H}^+$ itself provide a natural family of worldlines whose relative displacements can be used to define a notion of memory. Indeed, for the corresponding notion of memory at null infinity, the initially stationary timelike geodesics near $\mathcal{I}^+$ limit, as $r \to \infty$, to the null geodesic generators of $\mathcal{I}^+$ in the unphysical spacetime. Thus, using the null generators of $\mathcal{H}^+$ to define black hole memory is closely analogous to using timelike geodesics near $\mathcal{I}^+$ to define memory at null infinity, with the only significant difference being that the memory effect for black holes occurs at ``zeroth order" on $\mathcal{H}^+$, whereas the memory effect for null infinity does not affect the structure of $\mathcal{I}^+$ itself.

The infinitesimal displacement vector, $\xi^A$, between null geodesic generators of the horizon satisfies the geodesic deviation equation on $\mathcal{H}^+$
	\begin{equation}
	k^i \mathcal{D}_i (k^j \mathcal{D}_j \xi^A) = - {R_{iBj}}^A k^i k^j \xi^B
	\label{geodev}
	\end{equation}
Given an initial deviation vector $\xi^A_0$ in the early time stationary era, we may solve \cref{geodev} to obtain the deviation vector $\xi^A_1$ in the late time stationary era. We may then compare $\xi^A_1$ with the parallel transport of $\xi^A_0$. If they differ, then the null generators can be said to have undergone a ``permanent relative displacement.'' The final displacement $\xi^A_1$ will depend linearly on $\xi^A_0$, so we could define a ``memory tensor'' as the linear map that relates these quantities, in analogy with \cref{DispTensorDef}. However, if defined in this manner, the memory tensor would be a map from vectors in the past stationary era to vectors in the future stationary era. In order to obtain a map on vectors in the future stationary era, we would have to transport $\xi^A_0$ to the future era\footnote{This difficulty does not arise at $\mathcal{I}^+$ because $\xi^A$ does not vary at $O(1)$.} by some means (e.g., parallel transport). This would appear to make this approach to defining memory considerably less useful.

We believe that a more useful notion of a memory tensor for black holes can be obtained by pursuing 
analogy with \cref{MemTensor}. \Cref{MemTensor} holds in $4$ dimensions in any choice of Bondi coordinates \eqref{scricoord}, 
and, as previously noted, an analog of \cref{MemTensor} holds in arbitrary dimensions 
in any choice of comoving coordinates (see section IIIE of \cite{SatishWald}). A permanent change in the relative displacement of a family of geodesics would be reflected by a change in the components of the metric in coordinates 
comoving with these geodesics. Gaussian null coordinates are suitable ``comoving coordinates'' for the null generators of $\mathcal{H}^+$. Therefore, it seems natural to {\em define} the memory tensor for a black hole horizon to be
	\begin{equation}
	\Delta_{AB} = \frac{1}{2} \Big[\gamma_{AB}\big|_{v_1} - \gamma_{AB}\big|_{v_0}  \Big]
	\label{membh}
	\end{equation}
where, in this equation, $A$ and $B$ represent components 
with respect to (any) Gaussian null coordinates on $\mathcal{H}^+$. Here $v = v_0$ is in the early time stationary era while $v = v_1$ is in the late time stationary era. 

We adopt \cref{membh} as the definition of the memory tensor for black holes. Since
	\begin{equation}
	\lie_k \gamma_{AB} = 2\,K^{(k)}_{AB} = \frac{2}{(D-2)}\, \vartheta^{(k)} \gamma_{AB} + 2\,\sigma^{(k)}_{AB}
	\end{equation}
we have that 
	\begin{equation}
	\Delta_{AB} = \frac{2}{(D-2)}\intf{v_0}{v_1}\vartheta^{(k)} \gamma_{AB}\,\mathrm{d}v + 2\intf{v_0}{v_1}\sigma^{(k)}_{AB}\,\mathrm{d}v
	\label{MemTensorH}
	\end{equation}
where this equation holds in (any) Gaussian null coordinates on $\mathcal{H}^+$.
	
	\subsection{Horizon Memory from Supertranslations}
	\label{BHMemProp}

As discussed in \cref{memnullinf}, at null infinity during a stationary era, there exist ``good cuts'' of $\mathcal{I}^+$, which are unique up to translations. If the spacetime is stationary at early and late times, the early time and late time good cuts will 
be related by a BMS supertranslation, $u \mapsto u +T(x^A)$. Different choices of the good cuts in the early time and late time eras will affect $T$ only by a translation. Thus, the $\ell > 1$ part of the supertranslation $T$ is uniquely determined by the condition that it map an early time foliation by good cuts into a late time foliation by good cuts. The memory tensor at $\mathcal{I}^+$ is then given in terms of $T$ by \cref{GWMemasST}.

We showed in \cref{KHFol} that there are analogous notions of preferred foliations of a Killing horizon. In fact, we found two inequivalent such notions of ``preferred foliation" in \cref{KHFol}. We shall not attempt to choose between these here and, in the following, will simply assume that one of these notions has been chosen.

We assume that in the early time stationary era, the event horizon is well described by a nonextremal Killing horizon of surface gravity $\kappa_0$ with Killing field $\chi^a_0$, and that in the late time stationary era, it is well described by a nonextremal Killing horizon of surface gravity $\kappa_1$ with Killing field $\chi^a_1$. We introduce Gaussian null coordinates on all of $\mathcal{H}^+$ with $k^a$ chosen to be 
an affinely parametrized tangent to the null generators.
We seek a CFP supertranslation, $T$, of the form\footnote{Recall that, in this context, we denote the affine parameter of $k^a$ by $V$ instead of $v$.} 
	\begin{equation}
	T: (V,x^A) \mapsto \big(e^{\phi_1(x^A)}\,V + \phi_2(x^A),\, x^A\big)
	\label{cfpsuper2}
	\end{equation} 
(see \cref{cfpsuper} above) such that $T$ takes the early time preferred foliation into the late time preferred foliation. 

Since the early time foliation is mapped into itself by $\chi^a_0$ and the late time foliation is mapped into itself by $\chi^a_1$, a necessary condition
for $T$ to take the preferred foliations into each other is that $T$ map the early time horizon Killing field $\chi^a_0$ into a constant multiple of the late time horizon Killing field $\chi^a_1$
	\begin{equation}
T^* \chi_0^a = c\,\chi_1^a
	\end{equation} 
for some constant $c$. To see the consequences of this, we note that in the early time era, $ \chi_0^a$ takes the form
\begin{equation}
\chi_0^a = \kappa_0\inp{V - V_0(x^A)}k^a
\label{chi0}
\end{equation}
for some function $V_0(x^A)$, whereas in the late time era, $ \chi_1^a$ takes the form
\begin{equation}
\chi_1^a = \kappa_1\inp{V - V_1(x^A)}k^a
\label{chi1}
\end{equation}
for some function $V_1(x^A)$. We find that
\begin{align}
(T^*\chi_0)\big|_{(V, x^A)} 
&=  \kappa_0 \insb{e^{-\phi_1}\inp{V-\phi_2} - V_0} e^{\phi_1} k^a\\
&= \kappa_0\insb{V - e^{\phi_1}V_0 -\phi_2}k^a
\end{align}
Thus, a necessary condition for $T$ to take the preferred foliations into each other is 
\begin{equation}
\kappa_0\inp{V - e^{\phi_1}V_0 -\phi_2}= c\,\kappa_1\inp{V - V_1(x^A)}
\end{equation}
i.e. we must have $c = \kappa_0/\kappa_1$ and\footnote{Note that $V = V_0(x^A)$ corresponds to what would be the bifurcation surface $\mathcal{B}_0$ of the early time Killing horizon and $V = V_1(x^A)$ corresponds to what would be the bifurcation surface $\mathcal{B}_1$ of the late time Killing horizon, so \cref{phi21} simply states that $T$ maps $\mathcal{B}_0$ to $\mathcal{B}_1$.} 
\begin{equation}
\phi_2(x^A) =  V_1(x^A) - e^{\phi_1 (x^A)}V_0(x^A)
\label{phi21}
\end{equation}

If \cref{phi21} is satisfied, then $T$ will map the early time preferred foliation into the late time preferred foliation provided that it also maps a single cross-section, $\mathcal{S}_0$, of the early time preferred foliation into a cross-section, $\mathcal{S}_1$, of the late time preferred foliation. Let $F_0(x^A)$ and $F_1(x^A)$ be such that $\mathcal{S}_0$ is given by $V = F_0(x^A)$ and $\mathcal{S}_1$ is given by $V = F_1(x^A)$. The condition that $T$ map $\mathcal{S}_0$ into $\mathcal{S}_1$ is then
	\begin{equation}
	F_1(x^A) =  e^{\phi_1(x^A)}\,F_0(x^A) + \phi_2(x^A)
	\label{xsecmap}
	\end{equation} 

It is clear that \cref{phi21} and \cref{xsecmap} together uniquely determine $\phi_1$ and $\phi_2$, and thus the CFP supertranslation $T$. However, there is a remaining freedom in the choice of cross-section, $\mathcal{S}_1$, into which $\mathcal{S}_0$ is mapped. Instead of mapping $\mathcal{S}_0$ into $\mathcal{S}_1$, we could have mapped it into a cross-section $\mathcal{S}'_1$ differing from $\mathcal{S}_1$ by a Killing translation along $\chi^a_1$, i.e.,
	\begin{equation}
	F_1(x^A) - V_1(x^A) \mapsto C \left[ F_1(x^A) - V_1(x^A) \right]
	\label{F1var}
	\end{equation} 
for some constant $C$. This change in $F_1$ induces the change
	\begin{equation}
	\phi_1(x^A) \mapsto \phi_1(x^A) + {\rm const.}
	\label{phi1var}
	\end{equation}
together with the corresponding change in $\phi_2$ given by \cref{phi21}. This is closely analogous to the translation freedom in the choice of BMS supertranslation relating the good cuts of $\mathcal{I}^+$ at early and late times.

In summary, given a notion of ``preferred foliation'' in the early and late time eras---such as the uniform expansion foliation or the ABL foliation discussed in \cref{KHFol}---we have shown that the early and late time preferred foliations are related by a CFP supertranslation $T$ given by \cref{phi21} and \cref{xsecmap}. Furthermore, $T$ is unique up to \cref{phi1var} and the corresponding change in $\phi_2$. 
This equivalence class of $T$ thereby provides a notion of the ``supertranslation memory'' of a black hole.

	\section{Discussion} 
	\label{Disc}
We have seen in \cref{bhmemten} that a memory tensor, $\Delta_{AB}$, can be defined for black holes which characterizes the permanent change in the relative displacement of the null geodesic generators of the event horizon, $\mathcal{H}^+$. As explained there, this definition, \cref{membh}, of the black hole memory tensor is closely analogous to \cref{MemTensor} at $\mathcal{I}^+$. We have seen in \cref{BHMemProp} that a CFP supertranslation, $T$, characterizes the change in the early and late time preferred foliations of $\mathcal{H}^+$. The association of a CFP supertranslation to the change in the early and late time preferred foliations of $\mathcal{H}^+$ is closely analogous to the association of a BMS supertranslation to the change in the early and late time foliations of $\mathcal{I}^+$ by good cuts.

However, what appears to be entirely missing at $\mathcal{H}^+$ is a relationship analogous to \cref{GWMemasST} between the black hole memory tensor, $\Delta_{AB}$, of \cref{bhmemten} and the CFP supertranslation, $T$, of \cref{BHMemProp}. In addition, there does not appear to be any relationship analogous to \cref{intflux} between the integrated flux \cref{HSTFlux} associated with a CFP supertranslation and the black hole memory tensor, $\Delta_{AB}$.

The key reason for this difference can be traced back to the fact that the dynamical changes to a black hole occur on the horizon, $\mathcal{H}^+$, itself. For example, if an initially Schwarzschild black hole absorbs gravitational radiation and becomes a Kerr black hole at late times, in addition to the nontrivial changes to the geometry of $\mathcal{H}^+$ which occur during the dynamical evolution, there is a nontrivial permanent change in the horizon geometry between early and late times. No diffeomorphism can undo this permanent change and thereby make the late time Kerr black hole ``look like'' the initial Schwarzschild black hole. By contrast, gravitational radiation reaching $\mathcal{I}^+$ does not change the structure of $\mathcal{I}^+$ itself. Gravitational radiation first affects the geometry near $\mathcal{I}^+$ at radiative order $1/r^{(D/2 - 1)}$, and permanent changes near $\mathcal{I}^+$ between early and late times first occur only at Coulombic order, $1/r^{(D - 3)}$ \cite{SatishWald}. The permanent change in the metric at Coulombic order {\em can} be undone by a diffeomorphism \cite{StromingerHigher, SatishWald}, which, as we saw in \cref{memnullinf}, is a supertranslation in $D=4$ dimensions. This accounts for the close relationship between memory and supertranslations at $\mathcal{I}^+$. However, no such relationship appears to exist for black holes.

In summary, there is a very strong analogy between a black hole event horizon, $\mathcal{H}^+$, and future null infinity, $\mathcal{I}^+$. This analogy extends to the ability to define a memory tensor at $\mathcal{H}^+$ and the ability to assign 
supertranslations
which characterize the difference between early and late time preferred foliations of $\mathcal{H}^+$. However, many of the interesting interrelations between different facets of memory at $\mathcal{I}^+$---such as the relationship between the memory tensor and supertranslations---do not appear to extend to $\mathcal{H}^+$.

\bigskip

	\textbf{Acknowledgements} \vspace{2.5mm}
	
	We would like to thank Venkatesa Chandrasekaran, \'Eanna Flanagan, Marc Mars, Kartik Prabhu, Gautam Satishchandran, and Jonathan Sorce for helpful conversations. This research  was supported in part by NSF Grant No. PHY 18-04216 to the University of Chicago. AR was also supported by the NSF GRF Program under Grant No. DGE-1656518.
	
	\bibliographystyle{unsrt}
	\bibliography{fbhmrefs}
	
\end{document}